# Order From Chaos: A Reconsideration of Fundamental Principles

Paul J. Werbos[1]

**Abstract**

**Every so often, in science, new empirical results compel us to reconsider exactly what we really do know and do not know with mathematical certainty about fundamental principles. Startling but highly credible results from Greffet et al [1,2] and from Popovic's group [3] raise serious questions about the conventional understanding of the Second Law of thermodynamics. Prigogine, in his last years, argued that the patterns of order and life we see in the universe *would* have been possible, in equilibrium, in a universe of infinite lifetime, even without a Big Bang, because of field effects [4]. Huw Price [5,6] has provided compelling arguments to prefer the initial formulation of Hawkings' theory of the Big Bang—the version in which the arrow of time sometimes flows backwards in time, with profound implications for the foundations of quantum measurement. This paper will offer one possible unified view of these issues, and suggest that new technologies may in fact be possible in principle.**

## 1. Introduction

The purpose of this paper is to suggest a possible resolution to certain contradictions which have emerged in recent research across a wide variety of fields. The mathematical details of this picture certainly need to be filled in, but it is hoped that this paper will be a useful starting point for future research.

The immediate starting point for this work was a set of three independent pieces of work: (1) highly credible theoretical and empirical papers by Greffet et al [1,2], suggesting that about 10 percent of the infrared heat energy radiated by certain nanopatterned materials comes out as infrared radiation coherent to a few wavelengths and focused in its direction of propagation; (2) highly credible theoretical and empirical work by Popovic's group, showing how "totally incoherent ambient" microwave radiation in the 3-18 GHz band can be converted to useful electricity with 20 percent efficiency, using spiral rectennas [3]; (3) a proposal by Martin Margala of Rochester to adapt a new type of rectifier [7] to allow rectification of (heat) radiation in the 3-18 THz band, even without exploiting any temperature *difference* in the ambient environment.

These pieces of work seem to suggest that it should be possible to convert heat to electricity (free energy) on earth. But that suggestion constitutes gross heresy, in light of the conventional understanding of the Second Law of thermodynamics.

In actuality, the literature on the Second Law is very fragmented, and far less conclusive than is commonly believed. This paper will argue that it probably is possible, in principle, to convert ambient heat on earth to electricity after all. Furthermore, it will discuss some connections to the issues of quantum foundations and of the physical basis for life.

The next section will summarize the picture which appears to emerge when I try to reconcile a wide variety of contradictory sources. Section 3 will include some additional details for each of the points in section 2, in order.

## 2. Summary of a Proposed Reconciliation

**2a. Entropy**: the entropy of a state of the universe is essentially just the logarithm of the probability of that state, according to some equilibrium probability density of the physical state in question. For some dynamical systems and equilibrium probability densities, there is no statistical correlation at all between the state at point x and the state at point y, for x≠y; in such cases, the global entropy function is a "local function" (sum of a function of local quantities only), and entropy corresponds to perfect disorder. But many dynamical systems -- including our universe -- are not in that class.

**2b. Device Types:** Most conventional discussions of the Second Law prove that energy cannot be extracted from heat in equilibrium, by a certain class of devices. Many such discussions assume or "prove" that it is impossible to violate Carnot's Laws, which are actually more restrictive than the Second Law as normally interpreted; yet it is certainly well established that many devices, like fuel cells, are not "heat engines," and can in fact violate the limits set out by Carnot. Much of the literature on the Second Law focuses on limited classes of devices like gas dynamic systems, physical barriers to molecules, or cyclic disturbances to fields -- broader classes than heat engines, but narrower than what is possible in principle with fields.

**2c. General Results:** A more general or rigorous treatment of equilibrium statistics, based on classical Hamiltonian field theory *or on* quantum statistical thermodynamics, ends up with exactly the same operator expression for possible equilibria pdf (expressed as density matrices). This exact classical-quantum equivalence is itself of some interest. The allowable equilibria are a "local" invariant measure (which would indeed represent "disorder"!), multiplied by a canonical grand ensemble operator, which is essentially just a Boltzmann distribution in operator form, with other terms in the exponent to represent conserved quantities other than energy.

**2d. Greffet(1):** The possibility of nonlocality in this pdf depends on the Boltzmann term. For practical purposes, solid state devices may be viewed as a collection of ions in rigid locations, providing a kind of "designer" potential field $V(x)$, influencing electrons and photons. Greffet's work is a very straightforward illustration of how it is possible to craft a potential field which yields "nonlocal correlations" in the movement of electrons and photons. The theoretical analysis used by Greffet et al [2] in designing their nanopatterned SiC material is essentially the same mathematics one would use in analyzing the Boltzmann dynamics -- and is a solid theoretical basis for believing that systems in this class are not subject to the same locality limitations as we see in gaseous systems.

**2e. Greffet(2):** The violation of the "Second Law" conventional wisdom all takes place in the Greffet experiments and theory, the most thoroughly validated starting point for this paper. Once it is agreed that ambient heat can be turned (at 10 percent efficiency) into well-ordered, nonrandom focused radiation, the further details are essentially mechanical. In principle, for example, Greffet's materials actually cause temperature gradients which could be exploited to yield free energy; however, in practical terms it would make more sense to convert the radiation itself into electricity directly, using known technological principles which are most efficient for that kind of application. For specialists in Greffet's area, there is nothing surprising or shocking about how he does what he does. He uses well-established filtering principles; however, the larger *implications* of what he has done are still very important.

**2f. Popovic**: Popovic's claims are highly credible and well demonstrated for ambient microwave radiation; however, they need to be understood in the microwave context. In that context, it is routine to have antennas at room temperature which are able to receive radiation at frequencies much lower than room temperature (compared using the $kT=h\nu$ relation), because of the usual degree of coherence, which allows one to neglect quantum effects. Large scale rectennas, developed for purposes of power beaming, have achieved efficiencies on the order of 60 percent or more. For the relatively coherent IR coming out of Greffet's materials, it would be reasonable to expect that a spiral rectenna scaled to THz frequencies should be able to achieve 20 to 60 percent efficiency, if rectifiers could be found to enable a spiral rectenna scaled down by a factor of 1,000 from Popovic's in all spatial dimensions. In theory, if heat flux were on the order of 100 watts per square centimeter, this would imply that a 10cm-by-10cm slab of SiC, with a lower "Greffet" layer and an upper spiral nanorectenna layer, should be able to achieve 200-600 watts of cooling and power extraction, if properly mounted; however, the nanotechnology to demonstrate/test such a spiral nanorectenna is only now coming into reach, and has yet to demonstrate anything like the efficiencies achieved at lower frequencies.

**2g. Backwards Time**: If all of this actually should work, it is not inconceivable that a time-reversed version could also be developed -- but the theory behind such hopes is far less clear at this time.

# 3. More Detailed Discussion of Points 2a-2g.

One could easily write an entire book on any one of the points 2a through 2g. But that would defeat the goal here of trying to develop a more integrated picture. This paper will only provide a few highlights of what underlies these points.

## 3a. Entropy (2a): Basic Principles and Common Misconceptions

The topics of entropy, disorder and heat death have attracted great public interest – and, for that reason, have spawned large misunderstandings in the general public and in other areas of science. This section will review a few of the most basic general properties of entropy – and contrast them with common misconceptions about the subject.

Many people have been taught to believe the following: "Entropy is a measure of disorder. Mathematicians have proven that all possible dynamical systems follow the laws of entropy. These laws state that entropy must always increase until it reaches its maximum... whereupon the system is in a state of total disorder or 'heat death,' and entropy fluctuates about the maximum."

Years ago, I attended a plenary talk by Melanie Mitchell of the Santa Fe Institute, where she described a wide number of simulations of interesting complex dynamical systems, similar to what she discusses in [8]. One of the audience got up and said that what she described was inherently impossible. The systems she described were totally "closed" in a mathematical sense; the state at time t+1 was always a function of the state at time t, without any external variables coming into it. Yet as (simulated) time went by, the simulations displayed all kinds of interesting patterns of emergent order, without ever settling down to anything even remotely like a heat death. Of course, her simulations were completely valid; the problem was with the questioner's understanding of what entropy is about.

Even within the field of chemical thermodynamics, it was once believed [9] that open systems would be governed by a local "entropy production" quantity which would play the same role as local entropy in closed systems. Perhaps for that reason, many leading scientists were violently skeptical about the possibility of stable *chemical oscillations* a few decades ago – until, after many years of struggle, it was finally accepted that interesting patterns of oscillation really could persist, at least if there were an influx of energy.

In a similar way, in the field of nonlinear dynamical systems, it was once believed that emergent phenomena could only take on a few simple possibilities – like stable point attractors or diffuse "heat death" kinds of equilibria, like the extremes of fire and ice. Gleick has documented well the struggle [10] to have accepted the full range of possible equilibrium probability distributions, in effect. There is a great diversity of possible patterns of equilibrium statistics, depending on characteristics of specific dynamical systems.

Many would assume that complex behavior of this sort can only emerge in a certain kind of *open* systems, systems where energy can be exchanged with the environment. This would include systems which are mathematically closed, but which include mechanisms like forcing functions or dissipation terms which can change the energy level of the system. But it has been firmly established [11] that phenomena exactly like chaos can occur *within* "energy surfaces," even in systems which conserve energy in an absolute way. Classical or Poincare-style chaos is an important area for research. Back in 1994, I even found and simulated a simple set of stoichiometric equations which could conserve energy – and still yield oscillations in equilibrium, indefinitely [12] – *in the case* where one of the reactions is allowed to go "left to right" but not "right to left." (Of course, this was not a case of realistic chemistry, but a demonstration of a mathematical point.)

Despite these examples, there still do exist concepts of entropy which apply to very general classes of dynamical system. These generally come from the general formal theory of ergodic processes and related processes, developed initially by mathematicians like Von Neumann.

Many dynamical systems do possess one and only one possible equilibrium probability distribution function (pdf) for possible states. Crudely, if there are N possible states in a system, then the "entropy" of any state i is simply log $p_i$, where $p_i$ is the equilibrium probability of that state. More generally, the statement that "the actual probability distribution of the system converges to a maximum of entropy" is essentially the same as saying "the actual probability distribution converges to the equilibrium distribution," which is not all that surprising if there is only one possible equilibrium state (and we do time-averaging as we compute what we are converging to). This statement certainly does *not* say that we converge to a state of "heat death," As in Mitchell's simulations, it could simply mean that we converge to a set of interesting states. One may joke that "Washington D.C. has *already* reached a state of maximum entropy" – but this is not quite so much of a joke as it appears. The questions of order, life and "free energy" revolve about the *specific character* of the *specific entropy function* of the specific dynamical system under consideration, even for the case of closed systems.

Many of the strongest results about entropy were derived for a very specific class of dynamical systems – "stoichiometric" systems, systems of classical chemical reaction equations in which it is assumed that molecules move freely (with infinite mean free path) throughout infinite space, as in an ideal gas [9]. But this does not apply directly to systems like electromagnetic fields. Still, even in the stoichiometric case, it is well-known that we must complicate the analysis in a very important way: we must account for the fact that there exist *conserved quantities* like energy. Instead of having a unique equilibrium pdf, we have a *set* of possible equilibrium pdf, corresponding to different values of these conserved quantities. If we have n conserved quantities, $q_1 .. q_n$, the possible equilibrium pdf $p_i^*$ essentially take a generalized Boltzmann form:

$$p_i^* = p_i \exp(-c_1 q_1 - ... - c_n q_n)/Z \qquad (1)$$

where the constants $c_i$ are parameters of the distribution, where $p_i$ is a kind of default invariant probability measure (just a constant in the case of chemical systems), and where Z is a scalar inserted to make sure that probabilities add up to one across all states i. Z is called the "partition function" of the system. The concept of "temperature" becomes well-defined *after* we have such a distribution; if $q_1$ is chosen to be the energy of the system, then temperature can be *defined* as $a/c_1$, where "a" is a constant reflecting the arbitrary nature of the units in which we measure temperature.

Likewise, some dynamical systems can be described as systems made up of a finite or countably infinite number of independent objects i. *If* the states of these objects *happens* to be independent of the states of the other objects, then the global pdf takes on the form:

$$\Pr(\underline{s}) = \prod_{i=1}^{n} p_i(\underline{s}^{[i]}), \qquad (2)$$

where n is the number of objects (possibly infinity), where $\underline{s}^{[i]}$ is the state of object i, and $p_i$ is the equilibrium probability of object i being in that state. Taking logarithms, we may easily deduce that the global entropy function would be a "local" function (more precisely, a sum of local entropy functions):

$$S(\underline{s}) \equiv \log \Pr(\underline{s}) = \sum_{i=1}^{n} S_i(\underline{s}^{[i]}) \qquad (3)$$

It is a simple tautology to say that systems with equilibrium pdf like (2) would converge to a state of "disorder" (zero correlation between objects). Deducing (3) from (2) is not such a powerful result. The challenge is how to decide which systems would obey (2) in the first place.

The obvious next question is: how do we get closer to the actual universe we live in? Can we derive results which apply to *our* universe, which is certainly more complicated than these simple idealized approximations?

Section 3b will say more about that. But for now, let us discuss a class of dynamical systems which is much closer to our actual universe, permeated as it is by continuous fields. Let us consider systems whose state at time t is defined by a vector-valued field $\underline{\varphi}(\underline{x})$, for $\underline{x} \in R^3$. (I am not assuming a "vector" in the sense of relativity theory, but rather, any old mathematical vector, which could actually be some kind of amalgam of relativistic scalars, vectors, tensors, spinors, etc.). In this case, we can imagine *three* main possibilities for what the entropy function might look like:

(case 1): $\qquad S = \int S_{\underline{s}}(\underline{\varphi}(\underline{x})) d^3 \underline{x} \qquad (4)$

(case 2): $\qquad S = \int S_{\underline{s}}(\underline{\varphi}(\underline{x}), \nabla_{\underline{x}} \underline{\varphi}(\underline{x}), \nabla_{\underline{x}}^2 \underline{\varphi}(\underline{x})) d^3 \underline{x} \qquad (5)$

(case 3): $\qquad$ effectively nonlocal, not reducible to 5.

The key point is that case 1 implies a system which converges to total "disorder," just like equation 3. But case 2 does not. Even though case 2 involves a *kind* of local entropy function, it involves an entropy function whose minimum value need not be a kind of uniform "grey" field like a heat death! In summary, case 3 certainly would be exempt from the "disorder" results for simpler systems – but even case 2 can sometimes be exempt, depending on specific characteristics of the local entropy functions.

  We can find a simple example of case 2 under our nose, so to speak. In cosmology, it is perfectly obvious that our universe cannot possibly be converging to the heat death of the ideal gas, as one might imagine from the stoichiometric model. We are not converging to a situation where field states at nearby points are totally uncorrelated. John Wheeler [13] has developed a *modified* picture of the entropy death of the universe, as close as possible to the spirit of classical thermodynamics, but modified to account for the most obvious facts of life. In this picture, all life does die in the end... but the universe condenses out into objects like white dwarf stars, dead planets and black holes. These objects are not the kind of lively patterns we might wish for ... but they definitely *do* represent a kind of order, a kind of correlation across space in equilibrium for the proposed model of the universe. In my view, this order results from the *energy* term in the generalized Boltzmann distribution, a term which depends on field *fluxes* (as in equation 5), and which favors lower-energy states like states of gravitational condensation. In my view, the larger implications of the work of Greffet are probably based on the same kind of effect.

  Seth Lloyd, a leader in the field of quantum computing, wrote a paper in 1997 [14] arguing that field effects can allow a kind of "Maxwell's demon" in the quantum domain, in a way that may not depend so heavily on quantum effects as such. This has spawned a large literature, mainly focusing on quantum effects, which I have not yet had a chance to map out.

  By far the most rigorous general version of the Second Law in the current literature comes from the work of Allahverdyan. and Nieuwenhuizen [15]. Early versions of their work seemed to show promise in proving the general impossibility of extracting energy from heat, but the most recent work has demonstrated that counterexamples to the principles which lead to this can in fact be constructed. I have not yet assimilated the ideas of section 2 into their framework, but it would appear that the solid state field Hamiltonians and wave functions underlying Greffet's work do tie in with the kinds of theoretical loopholes they have described.

## 3b. Specific classes of "device" (2b): heat engines, fuel cells, organisms, etc.

It is extremely difficult to prove truly useful results for *all possible* universes, or even for *all possible devices* in our own universe. Thermodynamics has been forced to make severe simplifying assumptions or approximations, or to focus on limited classes of devices, in order to get practical results. The challenge before us is to improve the approximations, and increase the set of devices under analysis, without becoming mathematically intractable.

  Carnot's Laws are among the most useful results in thermodynamics. Decades ago, industry lobbyists in Washington routinely ridiculed conservationists who "do not understand how Carnot's Laws simply cannot be broken." Yet gradually the technical reality broke through these cultural barriers. Numerous authors explained how Carnot's Laws apply *only to heat engines*, and that fuel cells, *not being heat engines*, are exempt from Carnot's Laws (but not from the usual Second Law limitations). Of the hundreds of papers on this topic, I would prefer to cite Kordesch [16], because he – like Pat Grimes of Allis-Chalmers and Exxon, and John Appleby of Texas A&M – actually built alkaline fuel cell systems of great practical interest, showing efficiencies of about 60 percent without anything like the high temperature differences used in everyday car engines. Appleby and Kordesch both cite many more rigorous sources on the underlying thermodynamics of fuel cells. Papers on the "Second Law" which assert that all possible devices are bounded by Carnot's Laws are clearly relying on conventional wisdoms rather than general mathematics. (There is nothing wrong, of course, with papers which apply Carnot's Laws to those particular classes of device which are bound by them.) It is my view than carbon-tolerant alkaline fuel cells, along with "plug-in hybrids", are the two most promising technologies to permit drastic reduction in worldwide $CO_2$ emissions from cars soon enough to avert the total melting of the Arctic Ice Cap, if other measures are also taken [17].

  In chemical engineering, people have known for many decades how to decompose the *total* energy $\Delta H$ of a fuel into the free, extractable part $\Delta F$ versus the part $T\Delta S$ which only provides heat. But I have located no such generalized decomposition rule yet for the energy of ambient (unpolarized) electromagnetic fields, as a function of coherence length, frequency $\nu$, degree of uniformity in direction of motion, intensity, and temperature T of the receiver. (Formulas for signal detection do not apply directly, since energy extraction is a different issue.) Richard Fork [18] has reported an efficiency limit of $1-(kT/h\nu)$ for a specific class of optical switching systems; however, conventional rectennas clearly work with efficiency >0 for cases where $h\nu<<kT$!! Still, even Fork's formula, taken at face value, appears to suggest some possibility of extracting energy from those portions of the blackbody spectrum which exceed the average frequency.

If points 2e and 2f hold up under proper scrutiny, this would imply that we really cannot decompose ambient radiation into exploitable and nonexploitable parts, for all possible field devices. If people find it difficult to derive such a universal decomposition, this might explain why. (Even if we can only achieve a 1% extraction rate "per iteration," the other 99% of the heat energy remains available to exploit in a later iteration!) If so – instead of trying to decompose radiation into exploitable versus nonexploitable portions, we may have to limit ourselves to analyzing specific *classes* of field effect devices. In the worst case, theoretical limits could be developed separately for rectenna designs, and for light-modifying devices like Greffet's. This would provide more direct guidance as to what is possible along the lines of 2e and 2f. Of course, it would be desirable to make these classes as general as possible. These intellectual challenges will be important, one way or another, regardless of whether 2e and 2f hold up exactly as stated.

Some colleagues have suggested that the Greffet/nanorectenna combination could yield net energy, without violating today's understanding of the Second Law. After all, we would use this combination here on earth, which is an open system, receiving its heat directly or indirectly from the sun. But the concepts here really do not seem to involve the sun, so far as I can tell. As a thought experiment, they should work on an "infinite earth" if they work at all.

Finally, it is interesting to return to the age-old question of how we could better understand the phenomenon of life in terms of thermodynamics and nonlinear system dynamics. Life on earth certainly seems to obey the usual understanding of the Second Law, drawing its energy from the sun or from temperature differences at great depths; yet it should not surprise us too much that gelatinous cells based on DNA have not yet stumbled across strict nanopatterned SiC arrays!

The growing literature on (physics type) solitons in 3+1 dimensions [19-22] has some relevance both to thermodynamics and to the understanding of life. Black holes ala Wheeler are only one of the main examples of energy-minimizing systems which can be part of a "case 2" equilibrium state. "Solitons" and chaotic solitons are further examples, more suggestive of life and of interesting possible patterns. Many biologists have expressed great interest in modeling organisms or patterns within organisms, using the concept of soliton.

However – organisms clearly are not chaotic solitons. Chaotic solitons, as I have defined them, still retain the basic property of *absolute* stability. Biological organisms can be represented as ensembles of localized patterns, like chaotic solitons, but they do not possess absolute stability. There is an interesting analogy here: organisms are to chaoitons as optimal control is to robust control, or as economic self-sufficiency in space (exports equal imports) is to "total closed cycle habitats." Organisms age and die, to a very significant degree; their stability is purely a collective phenomenon, as birth rates exceed death rates under ideal conditions. Thus even when we try to abstract away from the specific conditions and forms of life we know of here on earth, it is hard to avoid the concepts of replication or autocatalysis even in the most abstract mathematical analysis. In fact, it is even hard to avoid some notion of "sexual" versus "asexual" replication, defined abstractly by whether the birth rate goes to zero in the limit as population density goes to zero. Yet some biologists have proposed cooperative notions of life which cannot be reduced to individual organisms in this way; again, there is an interesting analogy – to physics models which cannot be analyzed by using perturbation analysis around the usual elementary propagators.

## 3c. Equilibrium PDF (2c) for Classical and Quantum Fields – and Their Equivalence

I started out here by trying to disprove (if possible) a claim by Margala that electricity can be extracted from heat, using VLSI technology and field effects. Margala's proposed chip design was very different from the examples in section 2 – but I wanted to evaluate the general issue, in the general case. When I did not find a definitive proof for the general case, in a quick but intense search of the literature, I decided to do a quick derivation myself, based on methods I developed in [23-25]. For a wide variety of classical field theories, the entropy function turns out to fit "case 2" (equation 5). The classical PDF can be expressed equivalently in terms of "field operators," the mathematical concept which underlies the canonical version of quantum field theory (QFT) and quantum statistical thermodynamics (QST) [22, 23,26-30]. It turns out that this operator expression appears identical to the usual grand canonical ensemble in operator form used in QST. This equivalence between the statistics of classical fields and QST may have enormous theoretical implications, as I will discuss after sketching out the derivations. I will start out by reviewing the logic of well-known standard results – logic which is far more reliable than conventional wisdom as such.

All of the empirically supported theories of the physics of our universe – whether classical or quantum --- can be expressed in terms of Lagrangians and Lagrange-Euler equations. From the Lagrangian, one can work out the "Hamiltonian," the conserved energy density of the system.

For simplicity, let us start out by working out the statistics of a simple dynamical system, varying over time but not space, whose state $\underline{s}$ at any time is defined by knowing *two* real vectors, $\underline{\varphi}$ and $\underline{\pi}$, in $R^n$. Let us assume that the system is governed by a Hamiltonian function $H(\underline{\varphi},\underline{\pi})$. In this case, it is known that the dynamics of the system are:

$$\frac{d\varphi_j}{dt} = \frac{\partial H}{\partial \pi_j} \tag{6}$$

$$\frac{d\pi_j}{dt} = -\frac{\partial H}{\partial \varphi_j} \tag{7}$$

In order to work out the complete family of possible equilibrium pdf ($p(\underline{s})$), analogous to equation 1, we start out by deriving a simple member of that family, $d\mu(\underline{s})$, analogous to $p_i$ in equation 1. This kind of very simple equilibrium PDF is called an "invariant probability measure" in the literature on ergodic processes. However – it is not necessary at this stage that we scale $d\mu$ to make its integral add up to one; that comes later, when we divide by a scalar Z, as in equation 1.

I claim that the following probability measure is such an invariant measure:

$$d\mu(\underline{s}) = d\varphi_1 d\varphi_2 ... d\varphi_n d\pi_1 d\pi_2 ... d\pi_n \tag{8}$$

Intuitively, we might say that "the probability density $d\mu$ is constant or uniform," but equation 8 is more precise. To prove that $d\mu(\underline{s})$ is an invariant measure, we need only invoke the well-known Fokker-Planck equation which gives the dynamics of any probability distribution $\rho(\underline{s})$ over time:

$$\frac{\partial \rho}{\partial t} + \text{div}\, \rho v = 0 \tag{9}$$

where $\underline{v}$ refers to the "velocity" ($\partial_t \underline{s}$) of the system in the space of possible values of $\underline{s}$, and div is a standard vector function. In our case, we may write equation 9 more concretely as:

$$\frac{\partial \rho}{\partial t} + \sum_{i=1}^{2n} \frac{\partial}{\partial s_i}\left(\rho(\underline{s})\frac{\partial s_i}{\partial t}\right) = 0 \tag{10}$$

The condition for $\rho(\underline{s})$ to be an equilibrium is that $\partial_t \rho = 0$; thus the condition for equilibrium is that the big summation in equation 10 must be zero. To work this out further, we can break down the sum into parts, the part for states $s_i$ which are components of $\underline{\varphi}$, and the part for components in $\underline{\pi}$. This results in the equilibrium condition:

$$0 = \sum_{i=1}^{n} \frac{\partial}{\partial \varphi_i}\left(\rho(\underline{\varphi},\underline{\pi})\frac{\partial \varphi_i}{\partial t}\right) + \sum_{i=1}^{n} \frac{\partial}{\partial \pi_i}\left(\rho(\underline{\varphi},\underline{\pi})\frac{\partial \pi_i}{\partial t}\right)$$

$$= \sum_{i=1}^{n} \frac{\partial \rho}{\partial \varphi_i} \cdot \frac{\partial \varphi_i}{\partial t} + \rho \sum_{i=1}^{n} \frac{\partial}{\partial \varphi_i}\left(\frac{\partial \varphi_i}{\partial t}\right) + \sum_{i=1}^{n} \frac{\partial \rho}{\partial \pi_i} \cdot \frac{\partial \pi_i}{\partial t} + \rho \sum_{i=1}^{n} \frac{\partial}{\partial \pi_i}\left(\frac{\partial \pi_i}{\partial t}\right) \tag{11}$$

If we work out the second and fourth terms of this last expression, by substituting in from equations 6 and 7, we can see that those terms cancel out immediately, for *all* possible probability densities ρ(**s**), for a Hamiltonian system. For the particular choice of ρ(s)=dμ(**s**), the partial derivatives of ρ with respect to $\varphi_i$ and $\pi_i$ are all zero; this guarantees that the overall expression is zero, and that dμ(**s)** is an equilibrium (invariant) probability measure, as claimed.

In actuality, it is well-known in the Second Law literature [31] that simple Hamiltonian systems like this have this kind of simple equilibrium pdf. This insight is generally *applied* to the analysis of n distinct particles in a particle-based model. But we will take it further.

The distribution dμ(**s)** is only one member of a large family of possible equilibrium probability measures for this system. Let us assume that the system possesses exactly m conserved quantities, m functions $q_j$(**s)** which are guaranteed to remain constant over time. We know that m is at least 1, because the Hamiltonian itself is one conserved quantity. In this case, there is no interaction at all between the portions of the pdf which apply to different subspaces defined by different values of these conserved quantities. We may deduce that a*ny* pdf of the following form is also an equilibrium pdf:

$$p = f(q_1,...,q_m)d\mu / Z \tag{12}$$

where f is any nonnegative function, and where Z is the integral over all possible states **s** of fdμ; in other words, Z is that scaling factor which ensures that the integral of pdμ equals 1, as is required for a valid probability distribution. (In cases where Z is zero or does not exist, we do not have a valid pdf, except in cases where we can use the mathematical theory of tempered distributions to generalize the analysis.)

For a single system operating in complete isolation, there is no basis for requiring that $f(q_1,…,q_m)$ must take on the usual Boltzmann form as in equation 1. Conventional thermodynamics deduces this restriction by considering the case of *multiple* systems and making various types of assumptions about weak coupling between them. Perhaps those assumptions should also be revisited some day – but for now, it is interesting enough to ask what strange new technologies might be possible even within that restriction.

Next let us consider a more interesting case, the case of partial differential equations (PDE) over Minkowski space, still governed by Lagrange-Euler equations and so on.

A wide variety of interesting field equations can be represented by starting from a Hamiltonian functional H defined by:

$$H = \int \mathcal{H}(\boldsymbol{\varphi}(\mathbf{x}), \boldsymbol{\pi}(\mathbf{x}), \nabla\boldsymbol{\varphi}(\mathbf{x})) \, d^3\mathbf{x} \tag{13}$$

It is possible to *approximate* such a system by some kind of well-considered finite-element approximation, where the space of points **x** is represented by a discrete grid of points separated by some lattice spacing δ. Much of the original derivation of QFT was based on this kind of lattice-based approach [27,32,33]. As δ goes to zero, a well-constructed lattice approximation should converge in its predictions – and should converge to the continuous field theory. But now consider: if we have such a well-constructed finite approximation scheme, based on a Hamiltonian H(δ) and a lattice-based approximation $\mathcal{H}$(δ) to the Hamiltonian density, then at each stage of approximation, we still have a discrete Hamiltonian system, whose equilibrium pdf are given by equation 12! Thus if we accept the usual Boltzmann form for f, as discussed above, we arrive at a pdf for possible states of the fields defined by:

$$p = \exp(-c_1 q_1 - ... - c_m q_m) d\mu / Z \tag{14},$$

where we may write (roughly):

$$d\mu = \prod_{all\,\underline{x}} \prod_{i=1}^{n} d\varphi_i(\underline{x}) d\pi_i(\underline{x}) \tag{15}$$

A truly rigorous formulation of what equation 15 means would have to make use of the theory of functional integrals [22,33], but the common-sense meaning is reliable enough. For finite temperatures, the exponent in equation 14 keeps this all meaningful.

So far, we have essentially just confirmed that we are working with "case 2" as discussed in section 3a. This subsection has nothing more to say about extracting electricity from heat.

Equations 14 and 15 represent a family of pdf for the possible states S of the fields **φ**(**x**) and **π**(**x**) across space. They represent the pdf by explicitly stating the probability function or measure. However, there are other equally valid ways to specify or characterize a pdf. For example, probability distributions are often characterized by specifying the mean, the variance, the matrix of covariances, the skewness, the kurtosis, and all higher-order statistical moments. In [23], I proposed a new way to collect *all* the statistical moments of a set of classical fields into a single object, an operator in Fock-Hilbert space. (See [23,26] for tutorial introductions to operators in Fock-Hilbert space.) To conform with Weinberg [28], I considered the general case of classical fields over d-dimensional space; i.e., $\underline{x} \in R^d$. Equations 103-107 of [23] define the "classical density matrix" $\rho$ by:

$$\rho = \int \frac{\underline{v}(S)\underline{v}^H(S)}{|\underline{v}(S)|^2} \Pr(S) d^\infty S, \tag{16}$$

where:

$$\underline{v}(S) = \exp\left( c \sum_{j=1}^{n} \int (\theta_j(\underline{p}) + i\tau_j(\underline{p})) a_j^+(\underline{p}) d^d \underline{p} \right) |0\rangle, \tag{17}$$

$$\theta_j(\underline{p}) = \sqrt{w_j(\underline{p})} \int e^{-i\underline{p}\cdot\underline{y}} \varphi_j(\underline{y}) d^d \underline{y} \tag{18}$$

$$\tau_j(\underline{p}) = \frac{1}{\sqrt{w_j(\underline{p})}} \int e^{-i\underline{p}\cdot\underline{y}} \pi_j(\underline{y}) d^d \underline{y} \tag{19}$$

$$w_j(\underline{p}) = \sqrt{m_j^2 + |\underline{p}|^2}, \tag{20}$$

where "$\Pr(S)d^\infty S$" refers to the pdf whose statistical moments we are collecting, where $a_j^+$ is a standard creation operator (field operator), where $|0\rangle$ is the usual vacuum state of Fock-Hilbert space, and where the constant "c" is chosen to match the conventions in Weinberg. The classical density matrix turns out to be much easier to use and understand in the example where d=0 [23].

This particular way of collecting the statistical moments together has an extremely important property proven in [23]:

$$Tr(\rho g_n(\underline{\Phi},\underline{\Pi})) = <g(\underline{\varphi},\underline{\pi})>, \tag{21}$$

where "g" represents any analytic function of the fields, where $g_n$ is the normal form of the corresponding field operators, and where the angle brackets refer to the expectation value over the full ensemble of possible classical states of the classical fields. For example, when "g" is the Hamiltonian of the classical field, and $g_n$ is the (normal form) Hamiltonian operator, equation 21 gives us a way to calculate the expected energy of an ensemble of states. Furthermore, equation 16 is really just a way of averaging over possible states; for each possible state, equation 21 tells us that $g_n$ has the effect of multiplying $\rho$ by the function g, in effect.

Now I would like to introduce some rough conjectures which probably need cleaning up – but probably *can be* cleaned up to yield extremely important results.

First, I would conjecture that the invariant measure $d\mu$ of equation 15 corresponds to the (unnormalized) classical density matrix I (at least for some choice of the parameter "c".). A reason to expect this is that the off-diagonal terms of $\rho$ represent cross-correlations, while this $d\mu$ leads to zero cross-correlations across different points in space or different field components; also $d\mu$ and I are invariant with respect to translations in space and such.

Next, based on equation 21 and the properties noted in the paragraph which follows it, I would conjecture that the equilibrium pdf of equation 14 corresponds to classical density matrices $\rho$ of:

$$\rho = \exp(-c_1 Q_1 - \ldots - c_m Q_m)/Z \tag{22}$$

where Z is the usual scalar needed to normalize $\rho$, and where $Q_i$ is the normal form field operator representing the conserved quantity $q_i$. A reason to expect this is that equation 22 is actually the result of multiplying the invariant distribution $\rho$=I by a normal form operator. It might be necessary in principle to take great care with the possibility of noncommuting conserved quantities, or with making sure that this is not the *adjoint or dual* [25] of the equilibrium ρ; however, practical thermodynamics has not worked much with noncommuting conserved quantities, and the basic idea would still be workable in the adjoint representation.

The startling fact here is that equation 22 corresponds *exactly* to the family of density matrices used as the working foundation of quantum statistical thermodynamics (QST)! It is simply the usual grand canonical ensemble! The exact equivalence provides some additional justification for our previous, more classical discussion – but it may have larger implications.

Before discussing these implications, I should note that equation 21 – while new – is not unprecedented. For the special case of electromagnetic fields, similar properties have been known for a long time for mixtures of coherent states [29,34]. Walls and Milburn calculate a nonzero "variance" for coherent states – but the "variance" they compute is actually a second moment without subtraction of a mean. The coherent states they describe actually have a zero "variance" (as a statistician would define variance here), and they correspond to *definite* classical states, as per equation 16. The classic work of Coleman [34] and Mandelstam [35] on solitons in the quantum sine-Gordon field theory is all based on the *normal-form* Hamiltonian. The normal-form Hamiltonian is also the basis for all higher-order predictions of quantum electrodynamics. Elementary textbooks often stress the difference between Boltzmann statistics and Bose-Einstein statistics for individual photon modes, but the grand canonical ensemble operator of equation 22 is applied to density operators for the entire global system

Equation 22 appears to confirm Einstein's conjecture, long ago, that the predictions of quantum theory could be derived as predictions of the statistical dynamics of classical fields. This would suggest that classical field theory might itself be viable as a way of formulating the ultimate laws of physics. *If* the empirical predictions of quantum theory could all be derived as consequences of quantum statistical thermodynamics, equation 22 would imply that Einstein was right about this point as well.

In actuality, the situation is more complicated. There are two main complications – complications involving quantum measurement and complications involving fermionic fields.

First let us consider quantum measurement. The usual predictions of QFT are based on combining two kinds of calculation: (1) quantum dynamics, which ultimately boils down to predictions for bound stable states and predictions for "scattering states" [36] (i.e. equilibrium statistical states assuming a stable influx and outflux of particles; and (2) quantum measurement, usually handled by assuming the standard Copenhagen matrix projection formalism. The exact correspondence in equation 22 takes care of quantum dynamics – but what about quantum measurement?

In practical, empirical fields like quantum computing or quantum optics, the usual Copenhagen rules are no longer held to be exact in predicting what happens in "measurement objects" like polarizers and counters. For precise predictions [29,30], the measurement apparatus *itself* is represented as a solid state object governed by quantum dynamics. In effect, empirical work points us towards the classic Von Neumann/Wigner view of quantum measurement, in which true matrix projection only occurs "at infinity," and not in the finite universe of polarizers and brains that we live in. There is no empirical evidence of any "privileged objects" like brains being exempt from the usual quantum dynamics. I would conclude that empirical evidence now *compels* us to believe that the Copenhagen "laws" of measurement are really just approximations, whose partial validity is a consequence of quantum dynamics and cosmological boundary conditions. But if quantum measurement is a consequence of quantum dynamics and cosmological boundary conditions, then the same boundary conditions should imply the same measurement probabilities for classical fields as well, now that we know classical fields lead to the same predictions as "quantum dynamics."

Experts in quantum foundations would immediately ask how to reconcile these conclusions with Bell's Theorem experiments and similar paradoxes. Classical field theories like equation 13 are both "local" and "realistic." The Bell's Theorem experiments appear to rule out "local realistic" theories of physics. But in actuality, they rule out "local *causal* realistic" theories. There is good reason to believe that time-symmetric PDE would *not* imply the uniformly time-forwards flow of information at all levels of physics assumed in the name of "causality" in these theorems. [6, 23-25, 37]. Rather than ruling out PDE, these experiments rule out crude "common sense" thinking about time, the same kind of "common sense" which was once used to "disprove" special relativity. Einstein's comments about this kind of "common sense" remain as valid today as they were in his time. Section 3e will say more about this issue of information flows through time.

The careful reader may ask: how did we find an *exact* classical/quantum correspondence here, when we found (small) discrepancies before when comparing classical versus quantum equilibria in another context [24]? The obvious explanation is that we were studying sharper, closed sets of states before, whereas here we are studying thermodynamic ensembles. Before, we might have tried to "explain" Planck's constant by arguing that it goes away when we use "natural" units with $\hbar=c=1$. But now we are pushed into a more satisfactory new theory, in which Planck's constant represents a kind of temperature coefficient for the universe as a whole. This confirms an early speculation by Dirac that Planck's constant might be a kind of *emergent* quantity.

Next let us consider fermionic dynamics. The operator $a_j^+$ in equation 17 is a standard *bosonic* creation operator. Any bosonic quantum field theory whose Hamiltonian fits the form of equation 13 is equivalent to the corresponding classical field theory, in the statistical sense described above. But the standard model of physics [22] involves a *combination* of bosonic and fermionic fields. The predictions of quantum electrodynamics have been tested to great accuracy, in many types of experiment; classical field theories cannot match this experimental data, unless they can somehow reproduce fermionic statistics as well as bosonic. How could that be possible?

There are three obvious approaches to answering this question.

First, there is a mainstream way to answer it, which I do not like. Following Schwinger's source theory[38] or Feynman's functional integral approach[22], we may postulate "anticommuting classical fields." We may invoke the argument that such fields are somehow well-defined in terms of local representations of Clifford algebras. It would then be natural to augment equation 17 by adding terms which multiply such fields by fermionic creation operators. We would then end up with "classical field theories" in which some of the fields are ordinary continuous fields, while others are "classical anticommuting" fields. If we believe that such fields make mathematical sense, then we are home free. If not, we are not home free – but neither is Schwinger or Feynman.

Second, there is an approach suggested by soliton research [19,34,35] which still seems promising to me. One may first try to develop *bosonic* Lagrangians which replicate (or converge to) the standard model of physics. For example, Vachaspati [39,40] has laid out an exciting direction for research, based in part on 'tHooft's ideas about representing fermions as bound magnetic monopoles ('tHooft monopoles) or dyons. (Schwinger [41] also had suggestions for representing nucleons as bound systems of dyons, which may yet be worth revisiting.) The soliton-based approach would allow a classical field theory to reproduce all of the predictions of the standard model of physics – but it would also have two important further benefits. First, because of the classical/quantum equivalence, there is good reason to believe that these kinds of bosonic theories (with solitons) would be mathematically well-defined or "finite," *even without* regularization or renormalization; thus unlike the standard model, they could *explain* the masses of particles, instead of just hard-wiring the masses into the renormalization procedure. Second, in order to unify the standard model with general relativity, it would be good enough and simple enough to "metrify" the classical field theory, in exactly the same way that John Wheeler metrified Maxwell's Laws in 1956 in order to develop his "already unified field theory."

Based on the new results here, I am beginning to hope that we may have a third approach worth trying as well. It is still too early to know how promising it really is. The idea is that we may reconsider classical theories made up of a combination of continuous fields and point particles, as in the old Lorentz approach. We may add terms to equation 17 which multiply the "fields" for these point particles (fields which are actually Dirac delta functions!) by fermionic creation operators. I have explored this kind of idea in the past [42], because fermionic wave functions have a natural match to the statistics of point particles. I was not able to do anything useful with it then – but now, as an extension of equation 22 – it might provide us with a more elegant and direct and natural way to reproduce the predictions of the standard model exactly as it stands today. This would not avoid the need for regularization and renormalization, because even classical Lorentzian theories require renormalization. It may not be credible as an ultimate theory of physics. Nevertheless, it may give us the simplest, most direct way to explain all the experiments predicted by the standard model of physics. It may or may not be compatible with Wheeler's procedures for unifying classical field theories with gravity. There are some close relations between this approach and earlier work by Yasue and others [43], but here we are not assuming any random disturbance terms in the classical field theory.

## 3d. Greffet (2d,2e), Popovic (2f) and the Technology to Combine Them

Greffet's achievements are the key data which pushed me into the new heresy proposed in this paper. Yet Greffet's work, and Popovic's, stand on their own. There is little I can add to the citations and the discussion of sections 2d, 2e and 2f. It is all very straightforward, after we reconsider all the theoretical prejudices which might have blocked our ability to see the obvious.

Greffet's design was certainly based on crafting a potential field, V(**x**) (with consideration of other terms as well), and on predicting the behavior of his system by considering the eigenvectors of the resulting Hamiltonian. These eigenvectors, in turn, led to the interesting new wave functions and statistical correlations which are not spatially localized. There is absolutely no contradiction with the principles of quantum statistical thermodynamics as embodied in equation 22.

Some of my comments in sections 2d, 2e and 2f were based on questions I posed to leading specialists in these areas. I would be happy to acknowledge their help in greater detail, when and if they deem the time suitable. In many ways, I regret being compelled to publish a view so very far away from today's conventional wisdoms; yet, as some of those others have told me, the issues here are far too important to ignore or to judge n a hasty way. Of course, this paper is not intended as a kind of final judgment, even in my personal views. It is intended as a kind of starting point.

## 3e. Backwards Through Time?

Section 2g posed the question: could we possibly hope to create a *time-reversed* version of the kind of two-layer chip discussed in section 2f? To explain this question, we need to review a few basic issues regarding time.

Section 3c has given a condensed summary of the backwards time interpretation of quantum mechanics, which is discussed at length in [5,6,23-25,37,44], and many other papers. Many physicists who are not yet ready to return to time-symmetric PDE would still accept the idea that quantum measurement should be *derived* from quantum dynamics (e.g., a Schrodinger equation) and boundary conditions. But the Schrodinger equation is also time-symmetric in nature. The Copenhagen measurement rules are grossly asymmetric in time. Therefore, the only way to explain the asymmetry in measurement is to refer back to the time-asymmetry of the boundary conditions for our earth (and earth orbit, etc.), the place where we humans have done our physics experiments.

In [23], we proposed that the time-asymmetry in everyday quantum measurement is a consequence of the forwards arrow of time which mostly dominates our macroscopic life. (Probably Walls and Milburn[29] would say that this is not a speculation; for practical purposes, their book actually proves it.) That arrow of time, in turn, is mainly due to a straightforward boundary condition to our earth: a steady influx of ordinary free energy from the sun. Ordinary free energy is a kind of "improbable state" coming to us from the past which gives us the power to change the future. Yet logically, the time-symmetry of PDE and of quantum dynamics tells us that there formally exists a reversed-time mirror image of ordinary free energy. This may be called "backwards-time free energy." We cannot easily build devices to influence or communicate with the past , because we do not have access to a source of backwards-time free energy.

However… what if we really could build devices to convert heat to forwards-time free energy? Could we somehow time-reverse such designs, so as to extract backwards-time free energy instead? How would we even test such a possibility? If the answer to the first question should turn out to be "yes," then it would make sense to think hard about the two follow-on questions.

Even if we cannot extract free energy from heat, there are a number of other ways we might try to build technologies based on the backwards-time approach. For example, Huw Price has argued very effectively [4,5] that the original time-symmetric version of Hawking's theory of cosmology is more likely to be true than the later politically-corrected version. If Price is right, it should be possible to engineer real optoelectronic systems to detect (and even extract energy from) the backwards-time photons emitted from regions in the future where the arrow of time runs backwards. Even a tiny amount of backwards-time free energy might be interesting to work with. However, the technical details of such possibilities and of related quantum measurement experiments go well beyond the scope of this paper.

Sept.25,1968